      [1999/12/01 v1.4c Il Nuovo Cimento]
\documentclass{cimento}

\title{Evolutionary Quantization of Cosmological Models}
\author{Marco Valerio~Battisti\from{ins:x}\ETC,
Giovanni~Montani\from{ins:x}\from{ins:y}}
\instlist{\inst{ins:x} ICRA-International Center for Relativistic Astrophysics\\
Dipartimento di Fisica (G9), Universit\`a  di Roma, ``La Sapienza", Piazzale Aldo Moro 5, 00185 Rome, Italy.
  \inst{ins:y} ENEA-C.R. Frascati (Dipartimento F.P.N.), via Enrico Fermi 45, 00044 Frascati, Rome, Italy.}
\PACSes{PACS: {04.20.Jb}\hspace{0.3cm}{98.80.Bp-Cq}}
\begin{document}

\newcommand\be{\begin{equation}}
\newcommand\ee{\end{equation}}
\newcommand\bea{\begin{eqnarray}}
\newcommand\eea{\end{eqnarray}}
\newcommand\bseq{\begin{subequations}} 
\newcommand\eseq{\end{subequations}}
\newcommand\bcas{\begin{cases}}
\newcommand\ecas{\end{cases}}
\newcommand{\p}{\partial}
\newcommand{\f}{\frac}

\maketitle

\begin{abstract}
We consider a Schr\"odinger quantum dynamics for the gravitational field associated to a FRW spacetime and then we solve
the corresponding eigenvalue problem. We show that, from a phenomenological point of view, an Evolutionary Quantum Cosmology overlaps the Wheeler-DeWitt approach. We also show how a so peculiar solution can be inferred to describe the more interesting case of a generic cosmological model.
\end{abstract}

\section{Introduction}

In the search of a quantum theory of gravity one of the unsolved conceptual problems is the so-called {\it problem of time}; i.e. the absence of a time evolution for the wavefunctional in the framework of canonical quantization \cite{DeW67,Ku81,Ish92}. This feature is connected to the slicing procedure, in the sense that the slicing and the quantization can not be regarded as ``commutating'' operations. In fact, in the canonical approach, the construction of a quantum theory of gravity is based on the ($3+1$) slicing of spacetime (the ADM approach), i.e. the representation consisting of the unit time-like normal field and space-like hypersurfaces. Therefore it is possible move a criticism about the possibility to speak of an ADM procedure when we refer to a quantum spacetime; because, in this respect, either the time-like nature of vector field and either the space-like nature of the 3-hypersuprfaces can be considered only in the expectation value sense. In other words, it is rather ambiguous to apply the ($3+1$) slicing on a quantum level and therefore we have to recognize the impossibility of a physical slicing without frame fixing for a quantum spacetime \cite{Mo02,MeMo04a}. In fact a reference fluid has always a local light-cone structure, to be preserved in the quantum dynamics too.

In this work we apply an evolutionary quantum dynamics to a FRW cosmological model in the presence of a free inflaton field, the ultrarelativistic energy density of the thermal bath and a perfect gas contribution to account for Planck mass particle. Then, in order to discuss the phenomenological implication of the spectrum, we impose by hand, a cut-off length at the Planck scale. Finally, we will show that the solution for this symmetric model can be straightforward generalized for a generic inhomogeneous model.

\section{Evolutionary Quantum Gravity}

In this section we briefly analyze the implication of a Schr\"{o}dinger formulation of the quantum dynamics for the gravitational field \cite{Mo02,MeMo04a}. We require the states of the theory to evolve along the spacetime slicing so that $\Psi=\Psi(t,\left\{h_{ij}\right\})$; so the quantum evolution is governed by a smeared Schr\"{o}dinger equation which reads 
\be
i\p_t \Psi=\hat{\mathcal{H}}\Psi\equiv\int_{\Sigma_t^3}d^3x\left(N\hat{H}\right)\Psi
\ee
being $\hat{H}$ the super-Hamiltonian operator and $N$ the lapse function. If we now take the right expansion for the wave functional
\be
\Psi=\int D\epsilon\chi(\epsilon,\left\{h_{ij}\right\}) \exp\left\{-i\int_{t_0}^tdt'\int_{\Sigma_t^3}d^3x(N\epsilon)\right\},
\ee
the Schr\"{o}dinger dynamics is reduced to an eigenvalue problem of the form
\be\label{eipro}%
\hat{H}\chi=\epsilon\chi,\qquad \hat{H_i}\chi=0,
\ee
which outlines the appearance of a non zero super-Hamiltonian eigenvalue. Above we regard the super-momentum constraint as preserved, in order to deal with an actual $3$-space.

In order to understand the meaning of this super-Hamiltonian eigenvalue, we have to perform the classical limit of the above dynamical constraint. Therefore we replace the wave functional $\chi$ by its corresponding zero-order WKB approximation $\chi\sim e^{iS/\hbar}$. Under these restriction, the eigenvalue problem (\ref{eipro}) reduces to the classical counterpart, but characterized by the appearance of a new matter contribution, which admits the following energy density:  
\be\label{enden}%
\rho\equiv T_{00}=-\f{\epsilon(x^i)}{2\sqrt h}, \qquad h=\det h_{ij}.
\ee     
The explicit form of (\ref{enden}) is that of a dust fluid co-moving with the slicing 3-hypersurfaces, i.e. we deal with $T_{\mu\nu}=\rho n_\mu n_\nu$ ($n^\mu$ is the 4-velocity normal to the 3-hypersurfaces). We emphasize that this matter contribution has to emerge in any system which undergoes a classical limit and therefore it must concern the history of the Universe.

On the other hands it is possible to show that if we consider a gravitational system in the presence of a perfect fluid and adapting the spacetime slicing (i.e. looking the dynamics into the fluid frame) then we obtain the same relation (\ref{enden}) for the perfect fluid energy density. Therefore, a dualism between time and a frame of reference arises.

In the following we construct a solution in which we link the quantum number associated to the energy density of a perfect gas to the super-Hamiltonian eigenvalue $\epsilon$. In doing that we use the Planck mass particle (a perfect gas in the Planck era) as a ``clock'' for the quantum dynamics and, at the same level, we induce by them a non-relativistic matter component into the early Universe.

\section{Spectrum of the Quantum FRW Universe}

Now, we want to apply the Schr\"{o}dinger approach to a FRW Universe. The super-Hamiltonian of this model reads as ($R$ is the scale factor)
\be\label{eigenpro}%
H=-\kappa\f{p_R^2}R+\f{3}{8\pi}\f{p^2_\phi}{R^3}-\f{3} {4\kappa}R+R^3(\rho_{ur}+\rho_{pg}),
\ee
where $\kappa=8\pi l_P^2$. To reproduce the primordial Universe we have added to the dynamics of the system an ultrarelativistic energy density ($\rho_{ur}=\mu^2/R^4$), a perfect gas contribution ($\rho_{pg}=\sigma^2/R^5$) and a scalar field $\phi$ (a free inflaton field).

Performing the canonical quantization of this model we obtain the following eigenvalue problem (\ref{eipro}), with the right normal ordering \cite{Mo02}:
\be\label{H}%
\left[\kappa\p_R\f 1 R\p_R-\f{3}{8\pi R^3}\p_\phi^2 -\f{3} {4\kappa}R+R^3(\rho_{ur}+\rho_{pg})\right]\chi(R,\phi)= \epsilon\chi(R,\phi).
\ee 
The appropriate boundary condition for this problem are: i) $\chi(R=0,\phi)<\infty$ which relies on the idea that the quantum Universe is singularity-free and ii) $\chi(R\rightarrow\infty,\phi)=0$ that ensures a physical behavior at ``large'' scale factor.

The solution of the above eigenvalue problems reads as
\be
\chi(R,\phi)=\int\theta_k(R)\exp(i\sqrt{8\pi\kappa/3}k\phi)dk,
\ee
where 
\be
\theta_k(R)=\left(\sum_{n=0}^\infty c_nR^{n+\gamma}\right)\exp\left[-\f{\sqrt3} {4\kappa}\left(R+\f{2\kappa}3\epsilon\right)^2\right], \quad \gamma=1-\sqrt{1-k^2}, \quad c_0\neq0. 
\ee
The coefficients of the series obey the following recurrence relations
\be\label{rr}%
c_n=-f(n,\gamma)\left\{\left[-\f {2\epsilon} {\sqrt3}\left(n+\gamma-\f 3 2\right)+\f{\sigma^2}{\kappa}\right]c_{n-1}+\left[-\f {\sqrt3} {\kappa}(n+\gamma-2)+\f{\epsilon^2}3+\f{\mu^2}{\kappa}\right]c_{n-2}\right\},
\ee
with $f(n,\gamma)=((n+\gamma)(n+\gamma-2)+k^2)^{-1}$. Since we required the wave function to decay at large scale factor $R$ we have to terminate the series and therefore we obtain the spectrum of the Universe super-Hamiltonian and of the quantum number associated to the ultrarelativistic term, respectively:
\be\label{spee}%
\epsilon_{n,\gamma}=\f{\sqrt3\sigma^2}{2\kappa(n+\gamma-1/2)}
\ee
\be\label{mue}%
\sqrt3(n+\gamma)=\f{\kappa\epsilon^2_{n,\gamma}}{3}+\mu^2.
\ee
We emphasize that the ground state $n=0$ eigenvalue, for $\gamma<1/2$, is negative; therefore is associated via (\ref{enden}) to a positive dust energy density.

\section{Phenomenology of the Dust Fluid}

In order to analyze the cosmological implications of this new matter contribution, we have to impose a cut-off length in our model, requiring that the Planck length $l_P$ is the minimal physical length accessible by an observer ($l\geq l_P$). The existence of a fundamental Planckian lattice for the spacetime is expected in Quantum Gravity and it is a main result either in Loop Quantum Gravity \cite{RovSmo}, either in the String approach \cite{Pol}. So, from the thermodynamical relation for the perfect gas, we obtain a constraint on the $\rho_{pg}$ and then on the super-Hamiltonian eigenvalue:
\be
l^3 \equiv \f V {\cal{N}}=\f{3}{2}\f{l_P}{\rho_{pg}\lambda^2}\geq l_P^3 \qquad \Rightarrow \qquad \rho_{pg}\leq\mathcal{O}(1/l_P^4),
\ee      
where $l$ is the length per particle and $\lambda$ the corresponding thermal length ($\lambda=l_P$). Therefore we get $\sigma^2\leq\mathcal{O}(l_P)$ and so $|\epsilon_0|\leq\mathcal(1/l_P)$: {\it the spectrum is limited by below}. Another effect of the cut-off length is that the contribution of our dust fluid to the actual critical parameter of the Universe is provided by
\be
\Omega_{dust}\sim \f{\rho_{dust}}{\rho_{Today}}\sim\mathcal{O}\left( 10^{-60}\right).
\ee
Such a parameter is much less then unity and so no phenomenology can came out (today) from our dust fluid. In other words an Evolutionary Quantum Cosmology overlaps the Wheeler-DeWitt approach. Finally we face the question of the classical limit of the spectrum in the sense of large occupation numbers $n\rightarrow \infty$. As we can see from (\ref{spee}) the eigenvalue approaches zero as $1/n$. Therefore for very large $n$, our quantum dynamics would overlap the Wheeler-DeWitt approach.

We want to stress that, a priori, the new matter contribution that arise into the dynamics (i.e. a non relativistic fluid in the very early stages of the Universe evolution) can be regarded today as a good candidate for the so-called {\it cold dark matter} \cite{Ei,Ru}. However this is not the case, in fact the critical parameter of our dark matter candidate cannot fulfill unity in correspondence to an efficient inflationary scenario. As matter of fact is possible to see \cite{BaMo} that our dark matter candidate works only for an {\it e-folding} of about $23$, too small for a solution of the horizons paradox.

\section{The Generic Evolutionary Quantum Universe \cite{BaMo}}

In this last section we want discuss the more physical situation of a generic cosmological model, i.e. a model in which any specific symmetry has been removed \cite{BKL82}. In fact a quantum Universe has to be described by this kind of model because, roughly specking, in a quantum regime dealing with the absence of global symmetry is required by indeterminism; in fact on different causal regions the geometry `fluctuates' independently, so preventing global isometries. 

Although now we refer to a more complicated model then the previous one (the FRW model), nevertheless it is not difficult to show that the solution of the problem above has a more general feature that expected; in fact the super-Hamiltionian spectrum is, qualitatively, the same also for a generic quantum Universe.

The super-Hamiltonian of this generic model, adopting Misner-like variables $R$, $\beta_{\pm}$ \cite{Mis69a} ($R$ is the scale factor and $\beta_{\pm}$ describes the anisotropies), has the structure
\be\label{eigenpro}%
H(x^i)=\kappa \left[-\f{p_R^2}R+\f{1}{R^3}\left(p^2_+ +p^2_-\right)\right]+\f{3}{8\pi}\f{p^2_\phi}{R^3}-\f{R^3} {4\kappa l^2_{in}} V(\beta_\pm)+R^3(\rho_{ur}+\rho_{pg}),
\ee
where we added to the dynamics, as before, an ultrarelativistic energy density, a perfect gas contribution and a scalar field. Now we have to make same considerations about the above super-Hamiltonian:

i) It is well-known that classical behavior of a generic model is characterized by the fact that the spacial gradients can be neglected, and in this representation, each space point stands for a causal region (cosmological horizon). This feature can be easily see in a synchronous reference ($N=1$, $N^i=0$) where the dynamics reduces, point by point, to the one of a Bianchi IX model \cite{BeMo04}. 

ii) It is possible to see that a minimally coupled (classical) scalar field can suppress Mixmaster oscillations in the approach to the singularity of a generic cosmological spacetime \cite{Be00}. Then the dynamics reduces to a system of $\infty^3$ independent problems, in each space point isomorphic to a Bianchi I model. Therefore we can neglect, in the corresponding eigenvalue problem, the potential term for same $R^\ast\ll 1$.
     
iii) As soon as we neglect the potential term, the anisotropic contribution to the eigenfunction is isomorphic to the scalar field contribution. 

From these considerations we can conclude that either the different potential term either the anisotropies don't qualitatively modifies the super-Hamiltonian spectrum obtained above for the FRW model; in fact it comes out from the equation in $R$ which, apart from constants, retains the same form. Therefore we can apply locally the same phenomenological consideration also for a generic quantum Universe and then, also in this case, an Evolutionary Quantum Cosmology overlaps the Wheeler-DeWitt approach. Thus our approach can be inferred as appropriate to describe early stages of the Universe without significant traces on the later evolution.

\end{document}